# A possible way for getting the information out of a black hole

Rui Qi
Institute of Electronics, Chinese Academy of Sciences
17 Zhongguancun Rd., Beijing, China
E-mail: rg@mail.ie.ac.cn

We try to solve the problem about how to probe the inside of a black hole. We show that in the framework of revised quantum dynamics, which may naturally result from the combination of quantum mechanics and general relativity, the information inside a black hole can be gotten out in principle when using the conscious being as quantum measuring device.

## Introduction

As we know, although we can study the inside of a black hole using general relativity and quantum theory, present theories don't permit us to confirm their predictions there. The classical information can't be sent out from the inside of the black hole, or transmitted across its apparent horizon, while it is generally accepted that the Hawking radiation contains little information [1]. Even if some brave men enter into the black hole, and conduct the experiments to test present theories, they will not be able to send their results to us either. This is a bizarre situation, in which present theories don't permit to be confirmed inside a black hole. In fact, the situation is worse. If we can't confirm the predictions of our theories in principle, then our studies will essentially turn to be some metaphysical studies or even superstition.

In this paper, we will try to solve the above urgent problem. We show that the above unnatural situation may result from the incompleteness of present theories, which is also indicated by the well-known incompatibility between quantum mechanics and general relativity. We further argue that the incompatibility may imply the existence of dynamical collapse of wave function, whose theory is called after revised quantum dynamics [2-10]. In the framework of revised quantum dynamics, we find that the information inside a black hole can be gotten out in principle when using the conscious being as quantum measuring device. This provides a possible way to probe the inside of a black hole, and may solve the confirmation problem of our theories about the inside of a black hole.

## The possible existence of dynamical collapse of wave function

As we know, there exist some deep incompatibility between quantum mechanics and general relativity. In the domain of general relativity, the motion of particle and the space-time background are no longer independent, and there exists one kind of subtle dynamical connection between them. On the other hand, quantum mechanics must base on a fixed space-time background, which can't be influenced by the motion of particles. Thus the combination of these two theories may result in some essential incompatibility [8].

As a well-known example, we consider the superposition state of different positions, say position A and position B. On the one hand, according to quantum mechanics, the valid definition of this superposition requires the existence of a definite space-time structure, in which the position A and position B can be distinguished. On the other hand, according to general relativity, the space-time structure, including the distinguishability of the position A and position B, can't be predetermined, and it must be dynamically determined by the superposition state of particle. Since the different position states in the superposition state will generate different space-time structures, the space-time structure determined by the superposition state is indefinite. Then an essential logical incompatibility does appear.

Now let's see what can be inferred from the logical incompatibility. First, its appearance indicates that the superposition of different positions of particle can't exist when considering the influence of gravity, since the superposition state can't be consistently defined in principle. It should be stressed that this conclusion only relies on the validity of general relativity in the classical domain, and is irrelevant to its validity in the quantum domain. Thus the existence of gravity described by general relativity may result in the invalidity of the linear superposition principle, which will further result in the existence of dynamical collapse of wave function. Secondly, according to the definition of the superposition state of different positions of particle, its existence closely relates to the continuity of space-time. Thus the nonexistence of this superposition state implies that infinitesimal time interval based on continuous space-time will be replaced by finite time interval, and accordingly the space-time where the particles move will display some kind of discreteness. In fact, it proves that when considering both quantum mechanics and general relativity, the minimum measurable time and space size will be no longer infinitesimal, but finite Planck time and Planck length [11-13].

Lastly, we want to denote that the discreteness of space-time may also imply that the existence of dynamical collapse of wave function, and the many worlds theory [14] is not right. Since there exists a minimal time interval in discrete space-time, and each parallel world must solely occupy one minimal time interval at least, there must exist a maximal number of the parallel worlds during any finite time interval. Then when the number of possible worlds exceeds the maximal number, they will merge in some way, i.e. the whole wave function will dynamically collapse to a smaller state space. In recent years, many people have been trying to formulate the theory describing the dynamical collapse process of wave function, which is well known as revised quantum dynamics [2-10]. In such theoretical framework, the usual linear evolution equation of the wave function or Schroedinger equation is replaced by a stochastic linear or nonlinear equation. Presently, even if the last theory has not been found, but one thing is certain for the revised quantum dynamics, i.e. the collapse process does exist, and is one kind of dynamical process which will take a finite time interval to finish. Our following analysis will only rely on this common character of revised quantum dynamics.

## A possible way for getting the information out of a black hole

In this section, we will present a possible way for getting the information out of a black hole in the framework of revised quantum dynamics.

As we know, the information can't be sent out from the inside of the black hole by classical or local means. Thus we must resort to some kind of nonlocal means to get the information out of a black hole. As Bell's theorem and relevant experiments have indicated [15][16], there indeed exists nonlocal correlation or influence between the entangled particles or EPR correlating particles [17]. Then if we can transfer information by use of such nonlocal correlation, we may find the nonlocal means to get the information out of a black hole.

Unfortunately, present quantum theory do not allow us to transfer information nonlocally using such nonlocal correlation [18]. According to present quantum theory, even if there exists some nonlocal correlation or influence between the entangled particles, the measurement results on one of the particles are essentially random, and the information encoded in the correlation can't be obtained locally. The main obstacle is that the nonorthogonal single states can't be distinguished in the framework of present quantum theory. But according to the revised quantum dynamics, can the information obtained from measurement be much enough to distinguish the nonorthogonal single states? This is not a simple problem. In fact, no one has strictly demonstrated the impossibility of nonlocal information transfer in the framework of revised quantum dynamics so far. As we know, measurement is essentially an information-gathering process of the observer such as our human being or other conscious being, who wants to carry out such measurement, and how much information can be obtained through measurement is closely related to the theory. Thus it is possible that information can be nonlocally transferred in a different theory from present quantum theory. Now let's deeply analyze such possibility in the framework of revised quantum dynamics, according to which the dynamical collapse of wave function exists and continues for a period of time.

First, if the observer is involved in the measurement after the completion of the dynamical collapse process, he will evidently get no more information than in present quantum theory, thus no room is left for the possibility of nonlocal information transfer either. Once the dynamic collapse process is completed, projection postulate and Born rule will take effect, and the prediction of revised quantum dynamics will be the same as that of present quantum theory. Thus in order to obtain more information, which may help to distinguish the nonorthogonal single states and achieve nonlocal information transfer, the state of the observer should be entangled with the measured state before the completion of the dynamical collapse process. This is surely an unusual situation.

Secondly, if the state of the observer is entangled with the measured state before the completion of the dynamical collapse process, then whether can the observer get more information, and what is the added information? In the following, we will clearly show that, owing to the peculiar self-consciousness function of the observer, which is not possessed by usual physical measuring device, the observer may identify the intermediate process before the dynamical collapse process finishes, and the added information may help him distinguish the measured nonorthogonal single states. Then nonlocal information transfer is indeed possible in the framework of revised quantum dynamics.

Now we let the states to be distinguished are the following nonorthogonal single states $\psi_1$ and $\psi_1 + \psi_2$, and the initial perception state of the observer is $c_0$. Then after interaction the corresponding entangled state of the whole system is respectively $\psi_1 c_1$ and $\psi_1 c_1 + \psi_2 c_2$, where $c_1$ and $c_2$ is respectively the perception state of the observer for the states $\psi_1$ and $\psi_2$. We assume that the perception time of the observer for the definite state $\psi_1 c_1$, which is denoted by $t_P$, is shorter than the dynamical collapse time for the superposition state $y_1 c_1 + y_2 c_2$, which is denoted by $t_C$[1], and the time difference $\Delta t = t_C - t_P$ is large enough for the observer to identify. Then the observer can perceive the measured state $y_1$ or his own state $c_1$ after time interval $t_P$, while for the measured superposition state $y_1 + y_2$, only after the time interval $t_C$ can the observer perceive the collapse state $y_1$ or $y_2$, or his own corresponding state $c_1$ or $c_2$. Since the observer can also be conscious of the time difference between $t_P$ and $t_C$, he can easily distinguish the measured nonorthogonal single states $y_1$ and $y_1 + y_2$. Once the nonorthogonal single states can be distinguished, it is well known that the nonlocal information transfer can be directly achieved using the EPR correlation[2].

It can be seen that only the observer satisfies some condition can he distinguishes the measured nonorthogonal single states. We call such condition NIT (Nonlocal Information Transfer) condition, which requires that the perception time of the observer for the definite state is shorter than the dynamical collapse time for the perceived superposition state, and the time difference is large enough for the observer to identify. In the following, we will further demonstrate that the NIT condition is not irrational, and can be satisfied in essence, i.e. there should exist some kind of conscious beings satisfying the condition in Nature. First, the perception time of the conscious being may be mainly determined by the structure of his perception part, while the dynamical collapse time for the perceived superposition state is basically irrelevant to the concrete structure. Thus the perception time and dynamical collapse time are relatively independent. Then it is natural that the above NIT condition is satisfied for some kind of conscious beings, and for other conscious beings it is not satisfied. Secondly, with the natural evolution and selection the structure of the perception part of the conscious being will turn more and more complex, and the perception time will turn shorter and shorter. On the other hand, the perception part may turn smaller and smaller, and the energy involved for perception may turn less and less. Thus the dynamical collapse time will turn longer and longer according to most theories of revised quantum dynamics [2-10]. Then there will appear more conscious beings satisfying the NIT condition with the natural evolution and selection. In fact, owing to the availability of nonlocal information transfer, satisfying the NIT condition will be undoubtedly helpful for the existence and evolution of the conscious beings.

---

[1] It should be noted that, since the collapse time of a single superposition state is an essentially stochastic variable, which average value is $t_c$, we should consider the stochastic distribution of the collapse time in a strict sense, i.e. a small number of single states is needed for practical application. In the following discussions, we always simply take the collapse time as the average value $t_c$ unless state otherwise.

2 It should be noted that Squires also noticed the possible relation between explicit collapse and superluminal signaling from a slightly different point of views [19].

In one word, it is not irrational that, for some kind of conscious beings, the perception time for the definite state is shorter than the perception time or dynamical collapse time of the perceived superposition state, and the time difference is large enough for the conscious beings to identify. Thus even if our human being can not satisfy the NIT condition, other conscious beings may satisfy this condition. In fact, some evidences have indicated that our human being may satisfy the NIT condition [20-21].

## A suggested experiment for getting the information out of a black hole

In the following, we will further present a concrete experiment for getting the information out of a black hole based on the above method.

We first prepare a large number of EPR correlating particle pairs such as photon pairs in polarization-correlated state. Then we let a robot enter into a big black hole with all the one partner particle of each of these EPR correlating photon pairs. The other partner particles are kept in the lab outside the black hole[3].

Now we will show how the robot can send information out of the black hole to us. For simplicity we let the robot send the simplest Morse codes '0' and '1' to us. We assume if the robot wants to send the information code '0', it doesn't measure the polarization of the partner photons, and if the robot wants to send the information code '1', it measures the polarization of the partner photons in a stated direction. Besides, in the lab we set a two-channel polarizer, which allows the polarization components of photons both parallel to and perpendicular to the stated direction or transmission axis of the polarizer to be passed and spread along two different space directions. Then the separated photons are input from different directions to the sensitive part or eye of a conscious observer, who satisfies the above NIT condition. Now if the robot sends the code '1', the state of each EPR correlating photon pair will be collapsed due to the measurement of the robot. Then each partner photon in the lab will be in a definite state of polarization, and each will be input from a definite direction to the eye of the conscious observer. This will trigger a perception of the observer after a shorter perception time $t_P$; If the robot sends the code '0', the state of each EPR correlating photon pair will be not collapsed. Then each partner photon in the lab will be in a superposition state of polarization, and the photons will all be input from two different directions to the eye of the conscious observer. This will trigger a definite perception of the observer only after a longer collapse time $t_C$. Since the observer satisfies the NIT condition, he can differentiate these two situations and decode the Morse codes sent by the robot[4]. Thus we can indeed get the information out of a black hole using a conscious observer satisfying the NIT condition.

---

[3] Owing to the influence of gravity, the correlation between the EPR photons may be altered, but we can tune the experimental settings and measure the altered correlation in principle [22-23]. Especially when the experiment is conducted near a very big black hole, and the distance between the robot and the lab is not so far, the space-time curvature will be very small, thus the correlation can be kept in substance.

## Some further Discussions

Today, nearly all physicists believe that nonlocal information transfer or superluminal communication is impossible, since its existence is evidently inconsistent with special relativity. But we must notice the other side of the problem, i.e. special relativity may be partially wrong due to the existence of quantum nonlocality, and the absolute validity of the principle of relativity may be limited. In fact, it has been suggested that quantum nonlocality requires the existence of a preferred Lorentz frame [10][24-26]. Besides, the normal Big Bang solution of general relativity indeed provides a special Lorentz cosmos frame, in which the photons emitted from the Big Bang, namely the cosmos microwave background radiation is isotropic, and the temperature of radiation provides an absolute measure of the cosmos time. Thus it is not unreasonable to assume the existence of a preferred Lorentz frame, and so far as there exists a preferred Lorentz frame, nonlocal information transfer will not result in the usual causal loop and can exist in a consistent way. Moreover, the existence of the preferred Lorentz frame required by quantum nonlocality may further imply the availability of nonlocal information transfer [27].

On the other hand, even if the existence of a preferred Lorentz frame is accepted, most physicists may further argue that quantum mechanics also permits no existence of nonlocal information transfer [18]. But as we have shown, present quantum theory may be incomplete, and the incompatibility between quantum mechanics and general relativity may indicate the existence of dynamical collapse of wave function. Thus present quantum

---

4 Certainly, a proper synchronization between the robot and the observer is also needed for practical application. A simple way is to send the Morse codes periodically.

theory may not have the right to prohibit the existence of nonlocal information transfer either.

Then it is reasonable to consider the possibility of nonlocal information transfer in the framework of revised quantum dynamics. As we have demonstrated, even if the dynamical collapse process is an objective process, and consciousness can not change the collapse results, but its direct intervention can indeed get more information about the measured quantum state, which will help to distinguish the nonorthogonal single states and further achieve nonlocal information transfer. Thus it proves that revised quantum dynamics does permit the existence of nonlocal information transfer.

Now how the existence of nonlocal information transfer relates to the black hole? As we know, the black-hole information paradox, which was first found by Hawking [1], hasn't been completely solved so far. It is generally accepted that a resolution of the paradox will need a much deeper understanding of the interplay between general relativity and quantum mechanics, and may requires the revision of present unitary quantum mechanics [1][28]. Some physicists even suggested that a resolution of the paradox may require the existence of nonlocal information transfer [29]. Besides, the analogy between the black-hole information lost process and consciousness process is also been seriously discussed [30].

On the basis of these recent developments, the discussion of the relations between consciousness, nonlocal information transfer and the black-hole information may be not unnecessary and absurd, but even very urgent. It can be seen that, in the framework of revised quantum dynamics, information is lost from our universe during the dynamical collapse process, since a pure quantum state evolves a mixed state after this process. This is consistent with the original proposition of Hawking [1]. In fact, if the dynamical collapse process is a general natural process, it will happen everywhere in our universe, and certainly there is no exception in the black hole. On the other hand, as we have demonstrated, the existence of dynamical collapse process may further permit the conscious being to get the information out of a black hole using his consciousness function, which will enable the confirmation of our theory about the inside of a black hole. This may save the black hole theorists from becoming metaphysicians or theologists.

## Conclusions

We conclude that, when using conscious being as quantum measuring device, revised quantum dynamics permits nonlocal information transfer. This provides a possible way for getting the information out of a black hole.